\documentclass{article}
\usepackage{spconf,amsmath,epsfig}
\usepackage[figuresright]{rotating}
\usepackage{amssymb}
\usepackage{graphicx}
\usepackage{float}
\usepackage{amsmath}
\usepackage{subfigure}
\usepackage{rotating}
\usepackage{multirow}
\usepackage{array}
\usepackage{algorithm}
\usepackage{algorithmic}
\usepackage{url}

\title{Density controlled Divide-and-Rule Scheme for\\ Energy Efficient Routing in Wireless Sensor Networks}

\name{A. Ahmad$^{\ddag}$, K. Latif$^{\ddag}$, N. Javaid$^{\ddag}$, Z. A. Khan$^{\$}$, U. Qasim$^{\sharp}$}

\address{$^{\ddag}$COMSATS Institute of Information Technology, Islamabad, Pakistan. \\
        $^{\$}$Faculty of Engineering, Dalhousie University, Halifax, Canada.\\
        $^{\sharp}$University of Alberta, Alberta, Canada.}
\begin{document}
%
\maketitle
\begin{abstract}
Cluster based routing technique is most popular routing technique in Wireless Sensor Networks (WSNs). Due to varying need of WSN applications efficient energy utilization in routing protocols is still a potential area of research. In this research work we introduced a new energy efficient cluster based routing technique. In this technique we tried to overcome the problem of coverage hole and energy hole. In our technique we controlled these problems by introducing density controlled uniform distribution of nodes and fixing optimum number of Cluster Heads (CHs) in each round. Finally we verified our technique by experimental results of MATLAB simulations.
\end{abstract}
\section{Background}
Lots of research has been done to control energy utilization in the network layer of WSN. Clustering technique got great attention in resolving energy utilization issue. In clustering technique, data is first gathered and then forwarded to Base Station (BS). A uniform distribution of nodes and optimum number of CHs in each round helps to control load distribution in clustering technique which, ultimately utilizes energy efficiently.

As a second step controlling the number of clusters formed during network operation enhances the network stability and lifetime. A network is said to be stable if difference between first node died time and last node died time is minimum \cite{1}. Optimum number of CHs not only control load distribution but also uses energy efficiently.

One of the problem in clustering technique is the creation of energy holes. In random distribution of nodes CHs which are overloaded cause the creation of energy holes. In multi-hop data forwarding technique, nodes near the BS consume large energy. These area of nodes are also called hotspots \cite{1}. Energy depletes quickly in the hotspot areas of network.

W.R. Hienzelman, A.P. Chandrakasan and H. Balakrishnan proposed LEACH \cite{2}; one of the first clustering routing protocol for WSNs. According to LEACH algorithm, selection of CH for current round is probabilistic. Therefore, in this approach, of CH selection CHs formed are not uniformly distributed, which may cause existence of disconnected nodes.

LEACH-Centralized (LEACH-C) is an extension of LEACH which is proposed by Balakrishnan, Chandrakasan and Heinzelman \cite{3}. The plus point of this algorithm is that, BS makes sure that node with less energy does not become CH. However in large scale network, nodes far away from BS are unable to send their status to BS.

Multihop-LEACH protocol is proposed by Nauman Israr and Irfan Awan \cite{4}. Multihop-LEACH has mainly two modes of operations, i.e., multihop inter-cluster operation and multihop-intra cluster operation. In the leading operation nodes sense the environment and send their data to CH, this data is received by BS through a chain of CHs, while the lagging operation is performed in time out period. However, in both modes of operations CH is selected randomly. This agreement does not guaranty full area coverage of the entire network which, it is monitoring.


W. Jun, \emph{et al}. \cite{5} proposed LEACH-Selective Cluster (LEACH-SC). CH selection in LEACH-SC is like LEACH. The algorithm changes the theme of cluster formation in such a way that node finds CH closest to the mid point between itself and BS, then joins that cluster. However, number of CHs fluctuate as rounds proceed.

Localization problem is discussed in \cite{6} and \cite{7}. Authors divided the network area in to sub areas. The localization technique helps to improve the coverage hole problem.

In this research work we introduce a new clustering technique of routing layer communication. In Density controlled Divide-and-Rule (DDR), nodes are distributed uniformly in the network and randomly distributed in different segments of network; to control the density. In this way coverage hole problem can be avoided. Secondly in DDR, clusters formed are static and number of clusters remain fix during network operation. The number of clusters formed are near to optimum number. This helps in efficient energy utilization and uniform load distribution. Rest of the paper is organised as under.
\section {Proposed Protocol Design}
In this section, we introduce DDR. First we describe how the network area is logically divided into segments then we find out the energy consumption in  these segments.
\begin{table}[h]
\centering
\caption{Notations used in Mathematical Model}
\begin{tabular}{|p{1.4cm}|p{6.5cm}|}
\hline
Symbols & Meaning  \\
\hline
   $\rho$ & Node Density \\
\hline
   $\phi$ & Data aggregation energy per round \\
\hline
$I_s$ & Internal square \\
\hline
$M_s$ & Middle square \\
\hline
$O_s$ & Outer square \\
\hline
$S_n$ & $n^{th}$ square \\
\hline
$C_p$ & Center point of network field \\
\hline
   $E^{T_{x}}_{I_{s}}$ & Transmit energy of internal square \\
\hline
   $T_{l}$ & Top Left \\
\hline
   $T_{r}$ & Top Right \\
\hline
   $B_{l}$ & Bottom Left \\
\hline
   $B_{r}$ & Bottom Right \\
\hline
   $E^{T_{x}}_{M_{s}\_node}$ & Transmit energy of middle square nodes \\
\hline
   $E^{T_{x}}_{M_{s}/S}$ & Transmit energy of middle square nodes per segment\\
\hline
   $E^{T_{x}}_{M_{s}\_CH}$ & Transmit energy of CH of middle square \\
\hline
   $E^{T_{x}}_{M_{s}\_all\_CH}$ & Transmit energy of all CHs of middle square \\
\hline
   $E^{Rx}_{M_{s}\_CH}$ & Receive energy of CH of middle square \\
\hline
   $E^{Rx}_{M_{s}\_all\_CH}$ & Receive energy of all CHs of middle square \\
\hline
   $E^{Tot}_{M_{s}}$ & Total energy of all regions of middle square \\
\hline
   $E^{Tot}_{O_{s}}$ & Total energy of all regions of outer Square \\
\hline
\end{tabular}
\end{table}


\subsection{Cluster Formation }
In DDR, static clustering technique is used. Nodes are uniformly distributed in the network but, randomly deployed in the clusters. BS segmentize the network area. Each segment is known as cluster.
Traditionally CHs are selected on the basis of threshold and optimum number of CH in each round. Nodes then select their CH on the basis of minimum distance and  received signal strength. In DDR, entire network field is divided into small segments. Division of the network field is such that communication distance between node and CH, and between CH and BS is reduced.

Initially network area is divided into $n$ concentric squares. The value of $n$ is relevant to the distance $d$ between two concentric squares. We find out that the value of $d$ is such that, the clusters formed be balanced in terms of area and density. For a network area of 120m x 120m the value of $d=20$. The value of $n$ can be found from the following equation.
\begin{equation}
n = \frac {C_p(x)} {d}   
\end{equation}
As the network area increases the value of $n$ increases. The value of $d$ from the centre point to the $n^{th}$ square increases in multiple of square number. If we have $n$ number of concentric squares then, by using following equations  we can find the co-ordinates of $n^{th}$ square.

\begin{equation}
T_r(S_n) = (C_p(x) + d_n , C_p(y) + d_n),  
\end{equation}

\begin{equation}
B_r(S_n) = (C_p(x) + d_n , C_p(y) - d_n),  
\end{equation}

\begin{equation}
T_l(S_n) = (C_p(x) - d_n , C_p(y) + d_n),   
\end{equation}

\begin{equation}
B_l(S_n) = (C_p(x) - d_n , C_p(y) - d_n).  
\end{equation}
\begin{equation}
T_l(S_n) = (C_p(x) - d_n , C_p(y) + d_n),  
\end{equation}

\begin{figure}[!ht]         
\centering
\includegraphics[height=5cm,width=8cm]{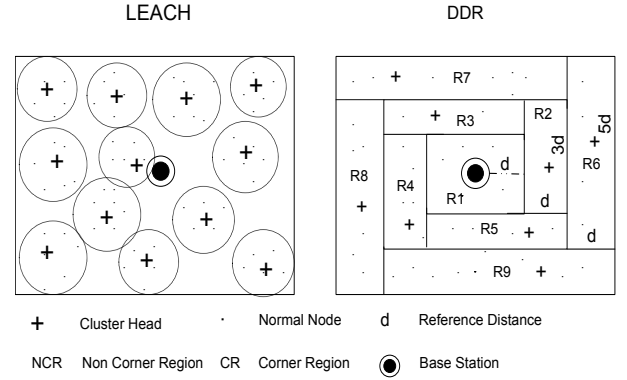}
\caption{Segment formation}
\label{DDRlayout}
\end{figure}

We placed the BS in the middle of the network field. As the area of $I_s$ is small therefore it is not further sub divided. The area between $I_s$ and $M_s$ is further sub divided into equal area rectangular segments using following equations. \\
Segment (S2)
\begin{equation}
B_l(S2) = B_l(I_s(x+d, y))
\end{equation}
\begin{equation}
T_r(S2) = T_l(I_s(x, y+d))
\end{equation}
Subtracting factor $d$ in the x co-ordinate of $T_r(I_s)$ and y co-ordinate of $B_r(I_s)$, we get divided the area into segments $S3$, $S4$ and $S5$. Similarly taking the corners of $M_s$ and adding or subtracting $d$ in x or y co-ordinates accordingly, we get divided the area between $M_s$ and $O_s$ into segments $S6$, $S7$, $S8$ and $S9$. Figure \ref{DDRlayout} shows the organization scheme of squares and segments.

\subsection{Energy Consumption in different Segments}
 This section describes mathematical model for energy consumption in the segments formulated in above section. Energy consumption in transmission or reception of a 1-bit packet over distance $D$ is obtained from \cite{8}.

Energy consumption in each segment of a square depends upon the number of nodes in a segment and energy consumed in transmitting fixed amount of data. Equations \ref{energyIs}, \ref{energyMs} and \ref{energyOs} gives energy consumption in $I_{s}$, $M_{s}$ and $O_{s}$ respectively.

\begin{equation} \label{energyIs}
  E^{T_{x}}_{I_{s}} = 4\rho d^{2} T_{Energy}
\end{equation}

\begin{equation}\label{energyMs}
  E^{T_{x}}_{M_{s}/Seg} = 4(3\rho  d^{2}-1)T_{Energy}
\end{equation}

\begin{equation}\label{energyOs}
  E^{T_{x}}_{O_{s}/Seg} = 4(3\rho  d^{2}-1)T_{Energy}
\end{equation}

\subsection{CH Selection and Energy Consumption of CH}
CH selection is an important step in any clustering protocol. In DDR, new CH is selected in each round in each segment. CH with minimum distance from central reference point is selected first then second least, third least and so on. Number of CHs in each round remain fix throughout network operation. In order to reduce communication distance multi-hop communication strategy is adopted in inter-cluster communication. CH in a segment selects front neighbouring CH as forwarding node.
CHs consume energy in transmission, receive and aggregation. Therefore, energy consumption is calculated individually. Area and node density of segment in same square is equal. There are total four CHs in four segments of a square. CHs in $M_s$ forward $O_s$ CH data too, therefore, CHs in $M_s$ expand more energy.
\begin{enumerate}
  \item {Transmit energy}
\begin{equation}
  E^{T_{x}}_{M_{s}\_all\_CH} = 4(3\rho  d^{2} + 4\rho  d^{2})T_{Energy} + 4\phi
\end{equation}
where $\phi$ is the data aggregation energy.

  \item {Receive Energy}
\begin{equation}
  E^{Rx}_{M_{s}\_all\_CH} = 12\rho  d^{2} -4
\end{equation}
\end{enumerate}
CHs in $O_s$ does not relay data. Therefore, CHs of $O_s$ expand energy in transmission and receive of local data.
\begin{enumerate}
  \item {Transmit energy}
\begin{equation}\label{Transmit Energy}
  E^{T_{x}}_{O_{s}\_CH} = (4\rho  d^{2})T_{Energy} + \phi
\end{equation}
\item {Receive Energy}
\begin{equation}
  E^{Rx}_{O_{s}\_CH} = (4\rho  d^{2} -1)R_{Energy}
\end{equation}
\end{enumerate}
%
%
%

\section{Performance Evaluation}        
In this section, we present the experimental results of our proposed protocol. The results are compared with existing protocols on the bases of four metrics, i.e., stability period, network lifetime, throughput and optimum number of CHs. Nodes are equipped with initial energy of 0.5J. Total nodes deployed in the network filed are 100 into an area of 100 x 100 $m^2$. BS is placed at the centre of the network field. Radio parameters are shown in table \ref{radio}.

\begin{table}[h]\label{radio}
\centering
\caption{Radio parameters}
\begin{tabular}{|p{4cm}|p{3.5cm}|}
\hline
Operation & Energy Dissipated  \\
\hline
Transmitter / Receiver Electronics  & Eelec=Etx=Erx=50nJ/bit \\
\hline
Data aggregation energy & EDA=5nJ/bit/signal \\
\hline
Transmit amplifier (if d to BS$<$do) & Efs=10pJ/bit/$4m^{2}$ \\
\hline
Transmit amplifier (if d to BS$>$do) & Emp=0.0013pJ/bit/$m^{4}$\\
\hline
\end{tabular}
\end{table}

\begin{figure}\label{allive}       
\centering
\includegraphics[height=7cm,width=9cm]{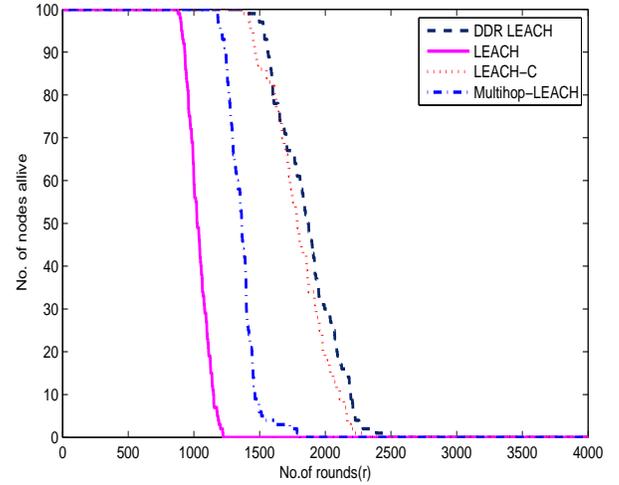}
\caption{Comparison: Alive Nodes}
\end{figure}

\begin{figure}[ht!]\label{throughput}       
\centering
\includegraphics[height=7cm,width=9cm]{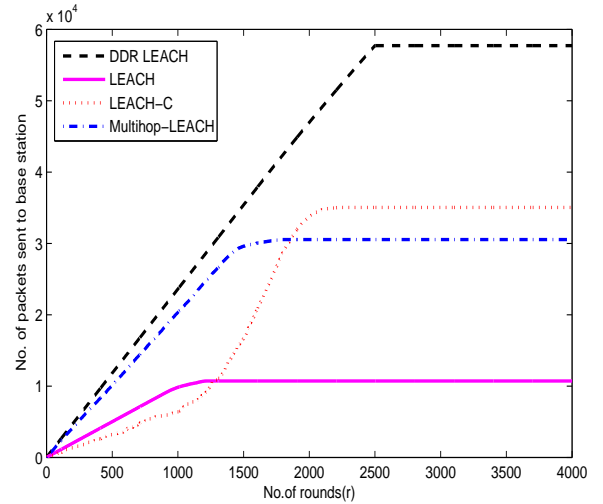}
\caption{Comparison: Throughput}
\end{figure}


In the MATLAB simulation environment all protocols are executed under same conditions. We have seen that DDR's stability period is 649 rounds better than LEACH and 148 rounds better than LEACH-C. DDR clustering approach minimizes communication distances and optimum number of CHs remain same in each round which, ultimately enhances stability period of the network.

Similarly in figure \ref{allive}, network lifetime of DDR network is 1270 rounds more than LEACH and 248 rounds more than LEACH-C. Balanced energy utilization and avoidance of coverage hole enhances the network lifetime.

Figure \ref{throughput} demonstrates that due to enhanced network life time and stability, packets sent to BS are 77.21$\%$ more than LEACH and 24.40 $\%$ more than LEACH-C.

 Table \ref{scalability} evaluates DDR by comparing it with LEACH-C and LEACH, in terms of network area and number of nodes. Table \ref{scalability} shows DDR is more scalable than others.
\begin{table}[H]\label{scalability}
\caption{Comparison of DDR, LEACH and LEACH-C for Scalability }
\centering
\begin{tabular}{|p{1.6cm}|p{2cm}|p{0.8cm}|p{0.9cm}|p{0.9cm}|}
\hline
Protocol& Network Area & No. of Nodes & First Node Death Round& Last Node Death Round\\
\hline
DDR  & 100m X 100m & 100 & 1496 & 2490 \\
&134m X 134m & 134 & 1460 & 2459\\
&150m X 150m & 150 & 1424 & 2388 \\
&200m X 200m & 200 & 1204 & 2270\\
\hline
LEACH-C &100m X 100m & 100 & 1348 & 2242 \\
&134m X 134m & 134 & 1189 & 2065\\
&150m X 150m & 150 & 1082 & 2059 \\
&200m X 200m & 200 & 1058 & 2123\\
\hline
LEACH & 100m X 100m & 100 & 847 & 1220 \\
&134m X 134m & 134 & 762 & 1100\\
&150m X 150m & 150 & 751 & 1093 \\
&200m X 200m & 200 & 746 & 1090\\
\hline
\end{tabular}
\end{table}
By increasing network area and number of nodes with a ratio of $1m^2$ per node; for 134 nodes we divide network area into four concentric squares,incase of 167 and 200 nodes we divide network area into five and six concentric squares respectively. In such situations communication distance is no more; less than or equal to reference distance and overhead on CHs near to BS increases as a result stability period and network lifetime decreases to an extent.

\section{Conclusion and Future Work}        
In this article we focused on energy efficient routing in WSNs. Our technique, DDR is based on static clustering and optimum number of CH selection in each round. In DDR we divided the network field into logical segments. The segmentation process helps to reduce communication distance between node and CH and between CH and BS. Multi-hop communication in inter-cluster further reduces communication distance. In DDR we have tried to overcome the problem of coverage hole and energy hole through density controlled uniform distribution of nodes in different segments of network. Optimum number of CHs in each round helps to achieve balanced load distribution. Which enhances stable period and network life time.



\end{document}